# A Rejoinder to Mackintosh and some Remarks on the Concept of General Intelligence

Moritz Heene

*Department of Psychology, Ludwig Maximilian University, Munich, Germany.*

Abstract

In 2000 Nicholas J. Mackintosh (2000) published an article in *Nature* referring to the concept of general intelligence ("*g*") claiming that there is clear empirical evidence for the existence of the *g* factor and psychologists are "…united in their support of *g*". Surprisingly, his view remained yet unchallenged although this issue is by no means as clear-cut as Mackintosh argues. Let us therefore attempt to clarify some common but unfortunately major misconceptions about *g*, which Mackintosh, following Jensen's (1998) precedent, recounted in his *Nature* article. The bottom line is that Spearman's *g* does not exist, that this has been known and acknowledged by leading scholars (Guttman, 1992; Thurstone, 1947) of factor analysis for decades so that the task of objectively defining human intelligence remains unfinished.

General intelligence, *"g"*, was first postulated by Charles Spearman in 1904 (Spearman, 1904). It has been debated for more than a century and is without a doubt one of the most important concepts of psychology. The crucial question is whether there really is a general intelligence in Spearman's sense, a single unidimensional trait, underlying all intelligence test performances. In *Nature* Mackintosh answered this question in the affirmative, concluding that: "…psychometricians were united in their support of *g*, or general intelligence" (Mackintosh, 2000) . Indeed, Jensen's (1998) redefinition of *g* which Mackintosh echoes as the first principal component of IQ test scores fulfils none of the stringent requirements of Spearman's *g* as we will see in the following.

Spearman postulated *one* general intelligence factor underlying the performances of all ability tests together with a specific factor unique to each test. The empirical consequences of his conceptualization of *g* are rather demanding, but this should not really come as a surprise, once one realizes how strong the necessary condition for *g* to exist actually is. In addition to positive test intercorrelations, the correlations matrix must also satisfy a second, very strong condition; namely that all tetrad differences (so-called off-diagonal second order minors) are zero. Then, in terms of these partial correlations,

Spearman's *g* only exists if all partial correlations (with *g* held constant) reduce statistically significant to zero. Numerous data sets exist which clearly violate this strong condition (Guttman, 1992).

Moreover, even if this double condition were met for any particular correlation matrix, this still would not confirm the existence of *g* because Spearman's conception of general intelligence was to explain the positive correlations among *all* intelligence tests, not just those of any particular battery of tests. Although initially there appeared to be some empirical support that Spearman's condition might be satisfied for small test batteries, his hypothesis of a general factor was soon falsified when larger test batteries were analyzed (Spearman, 1932). Even Spearman himself (1932) came to realize that *g* did not explain *all* correlations between tests. Instead, additional, so-called group factors (logical, mechanical, psychological and arithmetical abilities) were needed to depress the partial correlations to statistically acceptable small values – tantamount to the refutation of Spearman's *g*.

Mackintosh follows the precedent of Jensen (1998), asserting that *g* is equivalent to the first principal component in a factor analysis: "This 'positive manifold' means that factor analysis always yield a *large* general factor"(Mackintosh, 2000) (emphasis added). But this is always true for any first principal component (PC1); true in the sense that it is a mathematical tautology; but not true for the existence of *g*. The existence of any PC1, mathematically *defined* as the linear combination of the observed variables which has largest variance among all linear combinations (subject to the constraint that the weight vector has unit length) is tautologically true, and hence can not be an empirical discovery. Moreover, Spearman made no claims how "large" *g* is or should be. Rather, he claimed its existence must satisfy the tetrad difference equations. If one permits extraction of more than one principal components, then they (2, 3 or more components) will always account for the largest proportion of the variances of the observed variables, again by virtue of a mathematical tautology in the sense of explaining as much variance as possible. Therefore it is not legitimate for Mackintosh to infer "…the existence of the general factor *g*, extracted by factor analysis of any IQ-test battery"(Mackintosh, 2000) on the basis that the PC1 accounts for the largest proportion of the observed variance – it must. Mackintosh is correct only to the extent that a large general factor cannot be made to disappear because no statistical legerdemain can make a mathematical tautology disappear.

Jensen (1998) and his followers often argue that the identity of different PC1s extracted from different batteries are the same, even if they do not satisfy the tetrad difference condition, because the two regression weight vectors of the PC1's are usually very similar. This reasoning confuses random variables with distribution parameters. With the same logic one could argue that the variables are the same if their variances are the same. The equality of regression weight vectors tells absolutely nothing about equality of PC1s. Consider two sets of $p$ variables each, where all variables in the first set are perfectly uncorrelated with all variables in the second set, but both within-set correlation matrices are equal:

$$\begin{pmatrix} R & 0 \\ 0 & R \end{pmatrix}$$

If we extract the PC1 in each set, then we will obtain two identical columns of regression weights, but the PC1s of both batteries correlate zero.

Summarizing the critique in terms of common logic, let us see what went wrong in Mackintosh's argumentation: For each scientific proposition it is important to distinguish between hypotheses, tautologies, and meaningless statements. To illustrate these fundamental distinctions, consider the following three statements:

Theorem 1: The PC1 always explains the most of the variance due to its mathematical definition. This is well defined and true.

Theorem 2: There is single unidimensional trait $g$, underlying all intelligence tests performances so that the partial correlations are zero after holding $g$ constant. This is well defined and false.

Theorem 3: There is a linear combination of the original variables, called PC1. This is well defined and true, but tautological.

Theorem 4: There is an "…underlying cause of *g* (Mackintosh, 2000)". This is a meaningless statement as long as intelligence is empirically undefined.

In order to define intelligence objectively and to make his claim falsifiable Spearman stated a testable hypothesis which turned out to be false, i.e., theorem 2.

On the other hand, while trying to salvage a falsified theory by tendentiously redefining *g* as the first principal component (theorem 1), Mackintosh, among others, misrepresented a perfectly clear and falsifiable theory into various methods for data reduction with tautological conclusions (theorem 3), and discusses the characteristics of a still undefined variable (theorem 4). He talks as if he were dealing with theorem 2, thereby concluding that a false statement is true.

Finally, one is left with a misconception of a falsifiable hypothesis where a mere mathematical tautology, inherent in the structure of all positive test inter-correlations, is used to provide confirming "evidence".

Ironically, one of the main proponents of factorial intelligence theory, John Horn (2007), has late in life resigned himself to the plain fact that: "A wide array of evidence from research on development, education, neurology and genetics suggests that it is unlikely that a factor general to all abilities regarded as indicating intelligence produces individual differences variability. (…). These efforts have all failed to alter the conclusion that no general factor has been found. The evidence suggests that if there were such a factor, it would account for no more than a miniscule part of the variance in human intellectual abilities".

The logic of science tells us that before one describes and accounts the characteristics of a thing, one must show that it really exists. Otherwise it remains a myth (Shalizi, 2007).

6## Acknowledgements

I thank Prof. Trevor G. Bond, Hong Kong Institute of Education, for comments on an earlier draft.

Correspondence and Requests for materials should be addressed to: Dr. Moritz Heene, Unit Psychological Methodology and Assessment of the Department Psychology, Ludwig Maximilian University, Leopoldstr. 13, 80802 Munich, Germany, E-mail: heene@psy.lmu.de.
6